
\magnification 1200
\font\abs=cmr9

\def\uno{{\bf 1}}
\def\tens{\otimes}
\def\fraz#1#2{{\strut\displaystyle #1\over\displaystyle #2}}

\def\ket#1{|~#1~\rangle}

\def\ii#1{\item{$\phantom{1}[#1]~$}}
\def\jj#1{\item{$[#1]~$}}

\hsize= 15 truecm
\vsize= 22 truecm
\hoffset= 2.3 truecm
\voffset= 0.7 truecm

\baselineskip= 21 pt
\footline={\hss\tenrm\folio\hss} \pageno=1
\centerline
{\bf HEISENBERG XXZ MODEL AND }
\smallskip
\centerline
{\bf QUANTUM GALILEI GROUP.}

\bigskip\bigskip
\centerline{
{\it    F.Bonechi ${}^1$, E.Celeghini ${}^1$, R. Giachetti ${}^2$,
             E. Sorace ${}^1$ and M.Tarlini ${}^1$.}}
\bigskip

   ${}^1$Dipartimento di Fisica, Universit\`a di Firenze and INFN--Firenze,
\footnote{}{\hskip -.85truecm E--mail: GIACHETTI@FI.INFN.IT\hfill
   DFF 160/04/92 Firenze}

   ${}^2$Dipartimento di Matematica, Universit\`a di Bologna and INFN--Firenze.
\bigskip
\bigskip

{\bf Abstract.} {\abs The 1D Heisenberg spin chain with anisotropy of the
XXZ type is analyzed in terms of the symmetry given by the quantum Galilei
group $\Gamma_q(1)$. We show that the magnon excitations and the
$s=1/2$, $\ n$--magnon bound states are determined by the algebra. Thus
the $\Gamma_q(1)$ symmetry provides a description that naturally induces
the Bethe Ansatz. The recurrence relations determined by $\Gamma_q(1)$
permit to express the energy of the $n$--magnon bound states in a closed form
in terms of Tchebischeff polynomials.}
\smallskip
\noindent PACS 03.65.Fd; 75.10.Jm\hfill\break

\bigskip
\bigskip

Very recently some promising results have been obtained by the application
of the symmetry of inhomogeneous quantum groups [1] to physical systems with
a fundamental scale. In [2] the rotational spectra of heavy nuclei have
been reproduced, while in [3] and [4] applications to solid state problems
have been studied. In the former case the deformation parameter of the
quantum group is related to
the time scale of strong interactions, while in the latter a fundamental
length arises naturally from the lattice spacing. In [4] we have shown that
the symmetry described by the $q$--deformation of the Galilei group in one
dimension, $\Gamma_q(1)$, yields an algebraic scheme consistent with the Bethe
Ansatz [5,6] for solving the dynamics of quantum integrable models. The method
has been illustrated on the concrete example of the isotropic (or XXX)
Heisenberg ferromagnet. In this letter we shall use $\Gamma_q(1)$ for studying
a magnetic chain with anisotropy of XXZ type, whose properties have been
throughly investigated [7,8].
We shall also show that the conditions that are to be imposed to the mass of
the composite systems for obtaining bound states are connected with the
critical behaviour of the Casimir operator of $\Gamma_q(1)$.

The Hamiltonian of the model with periodic conditions
${\vec S_{N+1}}={\vec S_{1}}\ , ({\vec S}=S^x,S^y,S^z)$, is given by [7]:
$$ {\cal H}=2J\,\sum_{i=1}^N \Bigl(\,(1-\alpha)\,(S_i^x S_{i+1}^x +
               S_i^y S_{i+1}^y)
              + S_i^z S_{i+1}^z\,\Bigr)\ . \eqno(1)$$

Let $\ket{0}$ be the state with all the spin directed downwards. This is an
eigenstate of ${\cal H}$ with an energy given by $\epsilon_0 = 2J N s^2 $.
In terms of the states with one spin
deviate, $\psi=\sum\limits_i f_i\; S^{+}_i\ket{0}$, the eigenvalue equation
for ${\cal H}$ translates into the algebraic system
$$2Js\Bigl(\,(1-\alpha)\,(f_{i-1}+f_{i+1})-2f_{i}\,\Bigr) =
(\epsilon-\epsilon_0)f_{i}\ ,\eqno(2)$$
which leads to the dispersion relation
$$\epsilon-\epsilon_0= -4Js\Bigl(1-(1-\alpha)\cos k\Bigr)\ .\eqno(3)$$

We have shown in [4] that the isotropic analogue of system (2) can be described
by means of a quantum
group symmetry. Indeed, the solutions of (2) are obtained by evaluating the
solutions of the differential equation
$$-4Js\Bigl(1-(1-\alpha)\cos(-ia\partial_x)\Bigr)f(x)=
(\epsilon-\epsilon_0)f(x)\ \eqno(4)$$
at integer multiples of the lattice spacing $a$. This form of the Schr\"odinger
equation on the lattice has been proposed, e.g., in ref. [9].
For $a\rightarrow 0$, we recover from (4) the stationary
Schr\"odinger equation with an effective
mass $(-4Js\,(1-\alpha)\,a^2)^{-1}$ and with the symmetry of the 1--dim
Galilei group.
We were therefore led in [4] to introducing the deformation $\Gamma_q(1)$ of
the Galilei algebra, generated by the four elements $B,\ M,\ P,\ T\ $ with
commutation relations
$$ [B,P]=iM\ ,~~~~~~~~~~[B,T]=(i/a)\sin(aP)\ ,~~~~~~~~~~[P,T]=0\ ,$$
the generator $M$ being central.

The coproducts and the antipodes read
$$ \eqalign{
 \Delta B=e^{-iaP}\tens B+B\tens e^{iaP}\,,~~~~~~&~~~~~~
     \Delta M=e^{-iaP}\tens M+M\tens e^{iaP}\,,\cr
 \Delta P = \uno\tens P+P\tens\uno\,,~~~~~~&~~~~~~
 \Delta T = \uno\tens T+T\tens\uno\,,\cr
\gamma(B)=-B-aM\,,~~~~~\gamma(M)=-M\,,~&~~~~\gamma(P)=-P\,,~~~~~
\gamma(T)=-T\,,\cr}$$
while the Casimir of $\Gamma_q(1)$ is
$$C=MT-(1/a^2)\;(1-\cos(aP))\,.\eqno(5)$$

This quantum algebra admits the following realization in terms of differential
operators:
$$\eqalign{
B&=mx\,,~~~~~~~~~~M =m\,,~~~~~~~~~~P=-i\partial_x\,,\cr
T&=(m a^2)^{-1}\Bigl(1-\cos(-ia\partial_x)\Bigr)+c/m\,,\cr}$$
where $c$ is the constant value of the Casimir:
for $(m a^2)^{-1}=-4Js\,(1-\alpha)$ and $c/m=-4Js\alpha$,
the expression of $T$ coincides with the
operator on the left hand side of (4).

Like for the isotropic model  [4], the algebra is invariant under
$P\mapsto P + (2\pi/a)\,n$, the position operator is defined as $X=(1/M)\, B$
and the properties of the one magnon states are
obtained from the $\Gamma_q(1)$ symmetry.

The properties of the two--magnon states $\psi=\sum_{i>j} f_{ij}\;
S^{+}_iS^{+}_j\ket{0}$, with $f_{ij}=f_{ji}$, are then described by the
following system in the coefficients $f_{ij}$:
$$\eqalign{
\Bigl(\epsilon-\epsilon_0 + 8Js\Bigr)&\;f_{ij}
-2s(1-\alpha)\sum_n \Bigl(J_{nj}\,f_{in} + J_{in}\,f_{nj}\Bigr) \cr
{}&= - J_{ij}\Bigl((1-\alpha)\ (f_{ii} + f_{jj}) - f_{ij} - f_{ji}\Bigr)\ .
     \cr} \eqno(6)$$
The bonds $J_{ij}$ are equal to $J$ when the label $(ij)$ are nearest
neighbor pairs and vanish otherwise.
For $s=1/2$ the amplitudes $f_{ii}$ cancel in pair and $\Gamma_q(1)$,
which is a symmetry of the free system, permits a complete treatment of two
magnon excitations. Indeed, owing to the Bethe
Ansatz, which imposes the separate vanishing of the two sides of equation (6),
the interaction is reduced to
``boundary conditions'' ensuring that the homogeneous free
equation is satisfied at every pair of sites.

In our previous papers [2-4] we have already proved that the coproduct is
the correct operation which allows an algebraic treatment of the
many excitation systems. From $\Delta T$ we therefore find the following
two magnon energy $T_{12}$ :
$$\eqalign{
T_{12}&=T_1 + T_2 \cr
      {}&= (M_1 a^2)^{-1} \Bigl(1-\cos(aP_1)\Bigr) +
  (M_2 a^2)^{-1}\Bigl(1-\cos(aP_2)\Bigr)  + U_1 + U_2\ \cr}\eqno(7)$$
with $U_i=C_i/M_i\,,~i=1,2\,.$
For a fixed value of the spin $s$, the eigenvalue of $M_1=M_2=M$
is equal to $(-4Js\,(1-\alpha)\,a^2)^{-1}$ and $U_1=U_2=-4Js\alpha$.
Using the differential
realization $P_1=-i\partial_{x_1}$ and $P_2=-i\partial_{x_2}$,
the action of $T_{12}$ on the two magnon amplitude $f(x_1,x_2)$ reads

$$\eqalign{
T_{12}\,f(x_1,x_2)=&\,-8Js\,f(x_1,x_2)+2Js(1-\alpha)\,\Bigl(\,f(x_1,x_2+a)\cr
              {}& + f(x_1,x_2-a) + f(x_1+a,x_2) + f(x_1-a,x_2)\,\Bigr)\ .\cr}
\eqno(8)$$
The eigenvalue equation for $T_{12}$ is equivalent to the vanishing of the
left hand side of equation (6). Plane waves solve this eigenvalue equation
and the energy of the continuum is
$$\epsilon-\epsilon_0= -4Js\Bigl(2 - (1-\alpha)\cos(ap_1) -
(1-\alpha)\cos(ap_2)\Bigr)\ .$$
The two magnon state eigenfunctions are then obtained by imposing both the
periodicity and the Bethe
boundary conditions. We also observe that the coproduct of $P$ gives
$P_{12}=P_1+P_2$ for the total momentum, which, in the XXZ model,
is easily seen to be a conserved quantity.

Let us now show how the bound states for $s=1/2$ can be obtained from the
$\Gamma_q(1)$ symmetry.
We first observe that the central generator $M$ for the composite system reads
$$M_{12}=M_1 e^{iaP_2} + M_2 e^{-iaP_1}\ .$$
We then consider the energy for the two magnon system by rewriting (7)
in the form
$$T_{12}= (a^2\,M_{12})^{-1}\Bigl(1-\cos(aP_{12})\Bigr) + U_{12}\ ,\eqno(9)$$
where
$$U_{12}=U_1 + U_2 - \fraz{(M_{12} - M_1 - M_2)^2}{2\,a^2\,M_{12}\,
M_1\,M_2}\ .\eqno(10)$$
{}From equations (5), (9) and (10) the coproduct of the Casimir gives
$$C_{12}= M_{12}\,U_{12}\ .$$
The operators $C_{12}$ and $M_{12}$ label the irreducible representations of
the composite system and therefore must assume the same value over each state
of a given irreducible representation.
We shall now show that the critical behaviour of
$C_{12}$, as a function of the global mass $M_{12}$, defines
the Bethe conditions for the bound states and thus their energy. Indeed
${\partial C_{12}}/{\partial M_{12}}=0$ gives
$$ M_{12}=M_1+M_2+a^2M_1M_2(U_1+U_2)\, .\eqno(11)$$
In the continuum Galilei limit we find that $M_{12}$ is
identically equal to $M_1+M_2$, while, for $\alpha=0$
({\it i.e.} for $U_1=U_2=0$),
we recover the analogous condition for the isotropic XXX model [4].
For two $s=1/2$ magnons, in the above notations, equation (11) reads
$$M_{12}= 2M\,/\,(1-\alpha)\,.\eqno(12)$$
Defining $~2iv=P_1-P_2~$ the last condition yields just the Bethe Ansatz
for bound states [7]:
$$e^{-v}=(1-\alpha)\,\cos(aP/2)\,.$$
By substituting equation (12) into (9) and (10) we get the known form for the
energy of bound states:
$$T_{12}=-2J\Bigl(1-(1-\alpha)^2\,\cos^2(aP/2)\Bigr)\,.$$

We now give the generalization to the $n$--magnon case.
The total energy obtained from the quantum group can be written as
$$
T_{12\dots n}=\sum_{k=1}^n T_k =(a^2 M_{12\dots n})^{-1}
\Bigl(1-\cos(aP_{12\dots n})\Bigr) + U_{12\dots n}\eqno(13)$$
where $P_{12\dots n}=\sum_{k=1}^n P_k$ and
$$ U_{12\dots n}=
\sum_{k=1}^n U_k
  -\fraz 1{2a^2}~\sum_{k=2}^n \fraz{(M_{12\dots k} - M_{12\dots (k-1)}
- M_k)^2}{M_{12\dots k} M_{12\dots (k-1)}M_k}\,.\eqno(14)$$
In the above equations $M_{12\dots k}$ are defined by
iterating the coproduct and using the coassociativity:
$$M_{l\dots k} = M_{l\dots (h-1)}\ e^{ia(P_h+\dots+P_k)} +
M_{h\dots k}\ e^{-ia(P_l+\dots+P_{h-1})}\ ,\quad\quad l< h\leq k\ .\eqno(15)$$
Moreover the coproducts of the Casimir $C_{12\dots k}$ are found to be
$$C_{12\dots k} = M_{12\dots k}\,  U_{12\dots k}\,.$$

The bound states are obtained from (13) and (14) by imposing the vanishing
of the sequence of the derivatives of $C_{12\dots k}$ with respect to
$M_{12\dots k}$ for $k=2,\dots n\;$. The conditions determining $M_{12\dots k}$
read:
$$M_{12\dots k}=M_{12\dots (k-1)}+M_k +a^2\, M_{12\dots (k-1)}\,M_k\,
  (U_{12\dots (k-1)}+U_k)\;,\quad k=2,\dots n\;.\eqno(16)$$

We have performed a computer--assisted analysis [10] of the recurrence
relations
(14) and (16) and we have found that they can be solved, yielding

$$M_{12\dots k} =-\Bigl(2J(1-\alpha)a^2\Bigr)^{-1}\,{\cal
U}_{k-1}(1/(1-\alpha))
                              \ ,\quad k=2,\dots n\ ,\eqno(17)$$
$$T_{12\dots n}=\fraz{-2J(1-\alpha)}{{\cal U}_{n-1}(1/(1-\alpha))}\,\Bigl(
   {\cal T}_n(1/(1-\alpha)) - \cos(aP_{12\dots k})\Bigr) \eqno(18)$$

\noindent where ${\cal U}_k$ and ${\cal T}_k$ are the Tchebischeff
polynomials [11].
The equations (17), where $M_{12\dots k}$ have the form given in (15), are
equivalent to the Bethe conditions
$$\ M_{(k-1)k}=2M/(1-\alpha)=-\Bigl(J(1-\alpha)^2a^2\Bigr)^{-1}\,,~~~~~
               k=2\dots n\ .$$
Equation (18) gives the energy of the $n$--magnon bound states.

Some final remarks are in order. In the first place we observe that the
treatment given in [4] of the XXX model has been here extended in a
straightforward way also to the anisotropic XXZ model.
We thus support further evidence for the significance of
the application of the inhomogeneous quantum groups, like $\Gamma_q(1)$
and $E_q(1,1)$,
as kinematical symmetries of elementary physical systems described by a
discretized Schr\"odinger or Klein-Gordon equation:
it is the appropriate quantum group that
indicates the Bethe Ansatz and then the integrability of the system, so
that explicit computations are made possible.

\bigskip
\bigskip
\noindent{\bf Acknowledgments.}
The authors thank A.G. Izergin and V. Tognetti for useful discussions.

\vfill\break

\centerline{{\bf References.}}

\bigskip
\ii 1 E. Celeghini, R. Giachetti, E. Sorace and M. Tarlini, J. Math. Phys.
      {\bf 31}, 2548 (1990); J. Math. Phys. {\bf 32}, 1155 (1991);
      J. Math. Phys. {\bf 32}, 1159 (1991);\hfil\break
      E. Celeghini, R. Giachetti, E. Sorace and M. Tarlini,
      ``{\it Contractions of quantum groups}'', Proceedings of
      the First Semester on Quantum Groups, Eds. L.D. Faddeev
      and P.P. Kulish, Leningrad October 1990, Springer-Verlag, in press.
\smallskip
\ii 2 E. Celeghini, R. Giachetti, E. Sorace and M. Tarlini,
        ``{\it Quantum Groups of Motion and Rotational Spectra of Heavy
               Nuclei.}'', Phys. Lett. B (1992), in press.
\smallskip
\ii 3 F. Bonechi, E. Celeghini, R. Giachetti, E. Sorace and M. Tarlini,
        ``{\it Inhomogeneous Quantum Groups as Symmetry of Phonons.}'',
        University of Florence Preprint, DFF 152/12/91.
\smallskip
\ii 4 F. Bonechi, E. Celeghini, R. Giachetti, E. Sorace and M. Tarlini,
        ``{\it Quantum Galilei Group as Symmetry of Magnons.}'',
        University of Florence Preprint, DFF 156/3/92.
\smallskip
\ii 5 H.Bethe, Z.Phys. {\bf 71}, 205 (1931).
\smallskip
\ii 6 V.E. Korepin, A.G. Izergin and N.M. Bogoliubov, ``{\it Quantum Inverse
Scattering Method and Correlation Function. Algebraic Bethe Ansatz}",
(Cambridge, to appear in 1992).
\smallskip
\ii 7 R. Orbach, Phys. Rev. {\bf 112}, 309 (1958).
\smallskip
\ii 8 M. Takahashi and M. Suzuki, Progr. Theor. Phys. {\bf 48} 2187, (1972).
\smallskip
\ii 9 D.C. Mattis, Rev. Mod. Phys. {\bf 58} 361, (1986).
\smallskip
\jj{10} Wolfram Research, Inc., {\it Mathematica}, (Wolfram Research Inc.,
        Champaign, Ill., 1991).
\smallskip
\jj{11} M. Abramowitz and I.A. Stegun, {\it Handbook of Mathematical
Functions},
        (Dover Publications, Inc., New York, N.Y., 1972).
\bye